\documentclass[12pt]{article}

\usepackage{amssymb}
\usepackage{amsmath}
\usepackage{makecell}

\usepackage{graphicx}

\begin{document} %


\title{Probing anomalous quartic $\gamma\gamma\gamma\gamma$ couplings
in light-by-light collisions at the CLIC}

\author{
S.C. \.{I}nan\thanks{Electronic address: sceminan@cumhuriyet.tr}
\\
{\small Department of Physics, Sivas Cumhuriyet University, 58140,
Sivas, Turkey}
\\
{\small and}
\\
A.V. Kisselev\thanks{Electronic address:
alexandre.kisselev@ihep.ru} \\
{\small Division of Theoretical Physics, A.A. Logunov Institute for
High Energy Physics,}
\\
{\small NRC ``Kurchatov Institute'', 142281, Protvino, Russia}}

\date{}

\maketitle

\begin{abstract}
The anomalous quartic neutral couplings of the
$\gamma\gamma\gamma\gamma$ vertex in a polarized light-by-light
scattering of the Compton backscattered photons at the CLIC are
examined. Both differential and total cross sections are calculated
for $e^+e^-$ collision energies 1500 GeV and 3000 GeV. The helicity
of the initial electron beams is taken to be $\pm\,0.8$. The
unpolarized and SM cross sections for the same values of helicities
are also estimated. The 95\% C.L. exclusion limits on two anomalous
photon couplings $\zeta_1$  and $\zeta_2$ are calculated. The best
bounds on these couplings are found to be $6.85 \times 10^{-16}
\mathrm{\ GeV}^{-4}$ and $1.43 \times 10^{-15} \mathrm{\ GeV}^{-4}$,
respectively. The results are compared with the exclusion bounds
obtained previously for the LHC and HL-LHC. It is shown that the
light-by-light scattering at the CLIC, especially the polarized, has
a greater potential to search for the anomalous quartic neutral
couplings of the $\gamma\gamma\gamma\gamma$ vertex.
\end{abstract}

\maketitle


\section{Introduction} %

In the Standard Model (SM), the trilinear gauge couplings (TGCs)
\cite{Buchmuller:1986,Higawara:1987}  and quartic gauge couplings
(QGCs) \cite{Godfrey:1995}-\cite{Stirling:1999} are completely
defined by the non-Abelian $SU(2)_L \times U(1)_Y$ gauge symmetry.
These couplings have been accurately tested by experiments. A
possible deviation from the electroweak predictions can give us
important information on probable physics beyond the SM.

Anomalous TGCs and QGCs can be studied in a model independent way in
the framework of the effective field theory (EFT) via Lagrangian
\cite{Rujula:1992}-\cite{Degrande:2013}
\begin{equation}\label{eff_Lagrangian}
\mathcal{L}_{\mathrm{eff}} = \mathcal{L}_{\mathrm{SM}} +
\mathcal{L}_{(6)} + \mathcal{L}_{(8)} \;.
\end{equation}
The Lagrangian $\mathcal{L}_{(6)}$ contains dimension-6 operators.
It generates an anomalous contribution to the TGCs and QGCs. Let us
underline that the lowest dimension operators that modify the
quartic gauge interactions without exhibiting two or three weak
gauge boson vertices are dimension-8. The Lagrangian
$\mathcal{L}_{(8)}$ is a sum of dimension-8 genuine operators,
\begin{equation}\label{L_8}
\mathcal{L}_{(8)} =  \sum_i \frac{c_i}{\Lambda^4}
\,\mathcal{O}^{(8)}_i ,
\end{equation}
where $\Lambda$ is a mass-dimension scale associated with new
physics, and $c_i$ are dimensionless constants. This Lagrangian
induces anomalous deviation to the QGCs. It is assumed that the new
interaction respects the local $SU(2)_L \times U(1)_Y$ symmetry
which is broken spontaneously by the vacuum expectation value of the
Higgs field $\Phi$. CP invariance is also imposed. It means that
$\mathcal{L}_{(8)}$ is invariant under the full gauge symmetry. As a
result, the electroweak gauge bosons can appear in the operators
$\mathcal{O}^{(8)}_i$ only from covariant derivatives of the Higgs
doublet $D_\mu\Phi$ or from the field strengths $B_{\mu\nu}$,
$W_{\mu\nu}^a$.

There are three classes of dimension-8 operators. The first one
contains just $D_\mu\Phi$. It leads to non-standard quartic
couplings of massive vector bosons, $W^+W^-W^+W^-$, $W^+W^-ZZ$ and
$ZZZZ$. The second class contains two $D_\mu\Phi$ and two field
strength tensors. The third class has four field strength tensors
only. The dimension-8 operators of the last two classes induce the
anomalous quartic neutral couplings of the vertices
$\gamma\gamma\gamma\gamma$, $\gamma\gamma\gamma Z$, $\gamma\gamma
ZZ$, $\gamma ZZZ$, and $ZZZZ$. A complete list of dimension-8
operators which lead to anomalous quartic neutral gauge boson
couplings is presented in \cite{Eboli:2006}-\cite{Fichet:2014_1}. In
particular, the effective Lagrangian of the operators
$\mathcal{O}^{(8)}_i$ which contributes to the anomalous quartic
couplings of the vertex $\gamma\gamma\gamma\gamma$ looks like 
\begin{align}\label{Lagrangian_QNGCs}
\mathcal{L}_{\mathrm{QNGC}} & = \frac{c_8}{\Lambda^4} B_{\rho\sigma}
B^{\rho\sigma} B_{\mu\nu} B^{\mu\nu} + \frac{c_9}{\Lambda^4}
W_{\rho\sigma}^a W^{a\rho\sigma} W_{\mu\nu}^b W^{b\mu\nu} +
\frac{c_{10}}{\Lambda^4} W_{\rho\sigma}^a W^{b\rho\sigma}
W_{\mu\nu}^a W^{b\mu\nu} \nonumber \\
&+ \frac{c_{11}}{\Lambda^4} B_{\rho\sigma} B^{\rho\sigma}
W_{\mu\nu}^a W^{a\mu\nu} + \frac{c_{13}}{\Lambda^4} B_{\rho\sigma}
B^{\sigma\nu} B_{\nu\mu} B^{\mu\rho} + \frac{c_{14}}{\Lambda^4}
W_{\rho\sigma}^a W^{a\sigma\nu} W_{\nu\mu}^b W^{b\mu\rho} \nonumber \\
&+ \frac{c_{15}}{\Lambda^4} W_{\rho\sigma}^a W^{b\sigma\nu}
W_{\nu\mu}^a W^{b\mu\rho} + \frac{c_{16}}{\Lambda^4} B_{\rho\sigma}
B^{\sigma\nu} W_{\nu\mu}^a W^{a\mu\rho} \;,
\end{align}
see eq.~\eqref{zeta_c} below.

The explicit expression for dimension-8 Lagrangian in a broken phase
(in which it is expressed in terms of the physical fields $W^\pm$,
$Z$ and $F_{\mu\nu}$) can be found, for instance, in
\cite{Gupta:2012}.  We are interested in an effective Lagrangian for
the anomalous $\gamma\gamma\gamma\gamma$ couplings. It is given by
the formula \cite{Gupta:2012}
\begin{equation}\label{contact_Lagrangian}
\mathcal{L}^{\gamma\gamma\gamma\gamma}_{\mathrm{QNGC}} = \zeta_1
F_{\mu\nu}F^{\mu\nu}F_{\rho\sigma} F^{\rho\sigma} + \zeta_2
F_{\mu\nu}F^{\nu\rho}F_{\rho\sigma} F^{\sigma\mu} \;,
\end{equation}
where
\begin{align}\label{zeta_c}
\zeta_1 &= [ \,c_w^4 c_8 +s_w^4 c_9 + c_w^2 s_w^2 (c_{10} + c_{11})
\,]\Lambda^{-4} \;,
\nonumber \\
\zeta_2 &= [\,c_w^4 c_{13} +s_w^4 c_{14} + c_w^2 s_w^2 (c_{15} +
c_{16}) \,] \Lambda^{-4} \;.
\end{align}

The QGCs are actively studied for a long time. The anomalous $WWZZ$
vertex was probed at the LEP \cite{LEP_1} (see also
\cite{Belanger:2000}) and Tevatron \cite{Tevatron} colliders. The L3
Collaboration also searched for the $WWZZ$ couplings \cite{LEP_2}.
There have been investigations for the $WW\gamma\gamma$ couplings at
the LHC in \cite{Eboli:2001_1}-\cite{Zhu:2020}. The possibility of
measuring the $ZZ\gamma\gamma$ couplings were studied in
\cite{Eboli:2001_1}-\cite{Senol:2014}, \cite{Gupta:2012} and
\cite{Sahin:2012}. Recently, the LHC experimental bounds on QGCs
have been presented by the CMS \cite{CMS:QGCs} and ATLAS
\cite{ATLAS:QGCs} Collaborations. In a number of theoretical papers,
search limits for the anomalous vertex $WW\gamma\gamma$ at future
electron-proton colliders have been estimated
\cite{Ari:2020_1}-\cite{Rodriguez:2021}. The anomalous QGCs can be
also probed at linear $e^+e^-$ colliders \cite{Denner:2001}, in
particular, in the $e\gamma$ mode \cite{Eboli:1994,Atag:2007}
($WW\gamma\gamma$, $ZZ\gamma\gamma$ and $WWZ\gamma$ vertices) or
$\gamma\gamma$ mode \cite{Eboli:2001_2} ($WWWW$, $WWZZ$ and $ZZZZ$
vertices), \cite{Sahin:2009} ($WW\gamma\gamma$ and $ZZ\gamma\gamma$
vertices). Finally, in \cite{Koksal:2014,Koksal:2016} the anomalous
quartic couplings of the $ZZ\gamma\gamma $ vertex at the Compact
Liner Collider (CLIC) \cite{Braun:2008,Boland:2016} have been
examined. As one can see, in all these papers the anomalous QGCs
with the massive gauge bosons were examined.

The great potential of the CLIC in probing new physics is well-known
\cite{Dannheim:2012}-\cite{Franceschini:2020}. At the CLIC, it is
possible to investigate not only $e^+e^-$ scattering but also
$e\gamma$ and $\gamma\gamma$ collisions with real photons. In the
present paper, we will examine the possibility of searching for
anomalous $\gamma\gamma\gamma\gamma$ couplings in the light-by-light
(LBL) scattering with ingoing Compton backscattered (CB) photons at
the CLIC. Both unpolarized and polarized initial photons will be
considered. The first evidence of the process $\gamma\gamma
\rightarrow \gamma\gamma$ was observed by the ATLAS and CMS
Collaborations in high-energy ultra-peripheral PbPb collisions
\cite{ATLAS_ions, CMS_ions}. The LBL collisions at the LHC have been
studied in \cite{Atag:2010,Inan:2019}. Recently, the LBL scattering
at the CLIC induced by axion-like particles has been examined
\cite{Inan:2020_1,Inan:2020_2}.

\section{Light-by-light scattering in effective field theory} %

The $e^+e^-$ colliders may operate in $e\gamma$ and $\gamma\gamma$
modes \cite{Ginzburg:1981}. Hard real photon beams at the CLIC can
be generated by the laser Compton backscattering. When soft laser
photons collide with electron beams, a large flux of photons, with a
great amount of the parent electron energy, is produced. Let $E_0$
and $\lambda_0$ be the energy and helicity of the initial laser
photon beam, while $E_e$ and $\lambda_e$ be the energy and helicity
of the electron beam before CB. In our calculations, two sets of
these helicities, with opposite sign of $\lambda_e$, will be
considered, namely
\begin{align}\label{helicities}
(\lambda_e^{(1)}, \lambda_0^{(1)}; \lambda_e^{(2)}, \lambda_0^{(2)})
&= (0.8, 1; 0.8, 1) \;, \nonumber \\
(\lambda_e^{(1)}, \lambda_0^{(1)}; \lambda_e^{(2)}, \lambda_0^{(2)})
&= (-0.8, 1; -0.8, 1) \;,
\end{align}
where the superscripts 1 and 2 enumerate the beams. The helicity of
the photon with energy $E_\gamma$ obtained by the Compton
backscattering of the laser photons with helicity $\lambda_0$ off
the electron beam is given by the formula
\begin{eqnarray}
\xi(E_{\gamma},\lambda_{0}) = {{\lambda_{0}(1-2r) [1-x+1/(1-x)] +
\lambda_{e} r\zeta[1+(1-x)(1-2r)^{2}]}
\over{1-x+1/(1-x)-4r(1-r)-\lambda_{e}\lambda_{0}r\zeta (2r-1)(2-x)}}
\;,
\end{eqnarray}
where $x = E_{\gamma}/E_e$, $r = x/\zeta(1-x)$, $\zeta =
4E_eE_0/m_e^2$, $m_e$ being the electron mass.

The spectrum of the CB photons is defined by the helicities
$\lambda_0$, $\lambda_e$ and dimensionless variables $x$, $r$,
$\zeta$ as follows
\begin{align}
f_{\gamma/e}(x) = {{1}\over{g(\zeta)}} &\Big[ 1-x + \frac{1}{1-x} -
\frac{4x}{\zeta(1-x)} + \frac{4x^2}{\zeta^2(1-x)^{2}}
\nonumber \\
&+ \lambda_0\lambda_e r\zeta (1-2r)(2-x) \Big] ,
\label{photon_spectrum}
\end{align}
where
\begin{align}
g(\zeta) &= g_1(\zeta)+
\lambda_0\lambda_e \,g_2(\zeta) \;, \label{g_zeta} \\
g_1(\zeta) &= \left( 1 - \frac{4}{\zeta} - \frac{8}{\zeta^2} \right)
\ln{(\zeta + 1)} + \frac{1}{2} + \frac{8}{\zeta} - \frac{1}{2(\zeta + 1)^2} \;,
\label{g_zeta1} \\
g_2(\zeta) &= \left( 1 + \frac{2}{\zeta} \right) \ln{(\zeta + 1)} -
\frac{5}{2} + \frac{1}{\zeta + 1} - \frac{1}{2(\zeta + 1)^2} \;.
\label{g_zeta2}
\end{align}
The maximum possible value of $x$ is equal to
\begin{equation}\label{x_max}
x_{\max} = \frac{(E_\gamma)_{\max}}{E_e} = \frac{\zeta}{(1+\zeta)}
\;.
\end{equation}
The laser beam energy is chosen to maximize the backscattered photon
energy $E_\gamma$. This can be achieved if one puts $\zeta \simeq
4.8$, then $x_{\max} \simeq 0.83$.

\begin{figure}[htb]
\begin{center}
\includegraphics[scale=0.6]{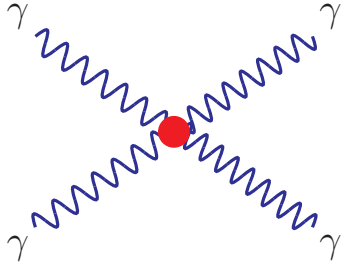}
\caption{The diphoton production in the collision of the
backscattered photons at the CLIC via anomalous quartic coupling.}
\label{fig:diagram}
\end{center}
\end{figure}

The LBL scattering of the CB photons happens as shown in
Fig.~\ref{fig:diagram}. Its differential cross section is expressed
in terms of the CB photon spectra, their helicities, and helicity
amplitudes \cite{Cakir:2008}
\begin{align}\label{diff_cs}
\frac{d\sigma}{d\cos \theta} &= \frac{1}{128\pi s} \int\limits_{x_{1
\min}}^{x_{\max}} \!\!\frac{dx_1}{x_1} \,f_{\gamma/e}(x_1)
\int\limits_{x_{2 \min}}^{x_{\max}}
\!\!\frac{dx_2}{x_2} \,f_{\gamma/e}(x_2) \nonumber \\
&\times \Big\{ \left[ 1 + \xi \left( E_\gamma^{(1)},\lambda_0^{(1)}
\right) \xi \left(
E_\gamma^{(2)},\lambda_0^{(2)} \right) \right] \nonumber \\
&\quad \times (|M_{++++}|^2 + |M_{++--}|^2) \nonumber \\
&\quad + \left[ 1 -\xi \left( E_\gamma^{(1)},\lambda_0^{(1)} \right)
\xi \left( E_\gamma^{(2)},\lambda_0^{(2)} \right)
\right] \nonumber \\
&\quad \times (|M_{+-+-}|^2 + |M_{+--+}|^2) \Big\} ,
\end{align}
where $x_1 = E_{\gamma}^{(1)}/E_e$ and $x_2 = E_{\gamma}^{(2)}/E_e$
are the energy fractions of the CB photon beams, $x_{1 \min} =
p_\bot^2/E_e^2$, $x_{2 \min} = p_\bot^2/(x_{1} E_e^2)$, $p_{\bot}$
is the transverse momentum of the outgoing photons. $\sqrt{s}$ is
the center of mass energy of the $e^+e^-$ collider, while $\sqrt{s
x_1 x_2}$ is the center of mass energy of the backscattered photons.
We will apply the cut on the rapidity of the final state photons
$|\eta_{\gamma\gamma}|<2.5$.

The physical potential of linear $e^+e^-$ colliders may be enhanced
if the polarized beams are used \cite{polarized_beams,CLIC_lum}. As
 will be seen below, it is exactly so in our case. For comparison,
similar results for unpolarized electron beams ($\lambda_e^{(1,2)} =
0$) will be also presented. Our calculations have shown that the
total cross sections are almost indistinguishable from the SM ones
for $\sqrt{s} = 380$ GeV (the first energy stage of the CLIC). That
is why, we will focus on the energies $\sqrt{s} = 1500$ GeV (the
second energy stage of the CLIC) and $\sqrt{s} = 3000$ GeV (the
third energy stage of the CLIC). The expected integrated
luminosities for these baseline CLIC energy stages \cite{CLIC_lum}
are presented in Tab.~\ref{tab:1}.
\begin{table}[h]
\centering
\begin{tabular}{||c|c||c|c|c||}
  \hline
\multicolumn{2}{||c||}{} & \multicolumn{3}{c||} {$L$, fb$^{-1}$}  \\
\hline
 Stage & $\sqrt{s}$, GeV & $\lambda_e = 0$ & $\lambda_e = -0.8$ & $\lambda_e = 0.8$ \\
 \hline
  2 & 1500 & 2500 & 2000 & 500 \\
  3 & 3000 & 5000 & 4000 & 1000 \\
  \hline
\end{tabular}
\caption{The CLIC energy stages and integrated luminosities for the
unpolarized and polarized initial electron beams.}
\label{tab:1}
\end{table}

We have calculated the differential cross sections
$d\sigma/dm_{\gamma\gamma}$, where $m_{\gamma\gamma}$ is the
invariant mass of the outgoing photons. Each of the amplitudes is a
sum of the anomaly and SM terms,
\begin{equation}\label{M_tot}
M = M_{\mathrm{anom}} + M_{\mathrm{SM}} \;.
\end{equation}
As the SM background, we have taken into account both $W$-loop and
fermion-loop contributions
\begin{equation}\label{M_ew}
M_{\mathrm{SM}} = M_f + M_W \;.
\end{equation}
The explicit analytical expressions for the SM helicity amplitudes
in the right-hand side of eq.~\eqref{diff_cs}, both for the fermion
and $W$-boson terms, are too long. That is why we do not present
them here. They can be found in [46].

\begin{figure}[htb]
\begin{center}
\includegraphics[scale=0.6]{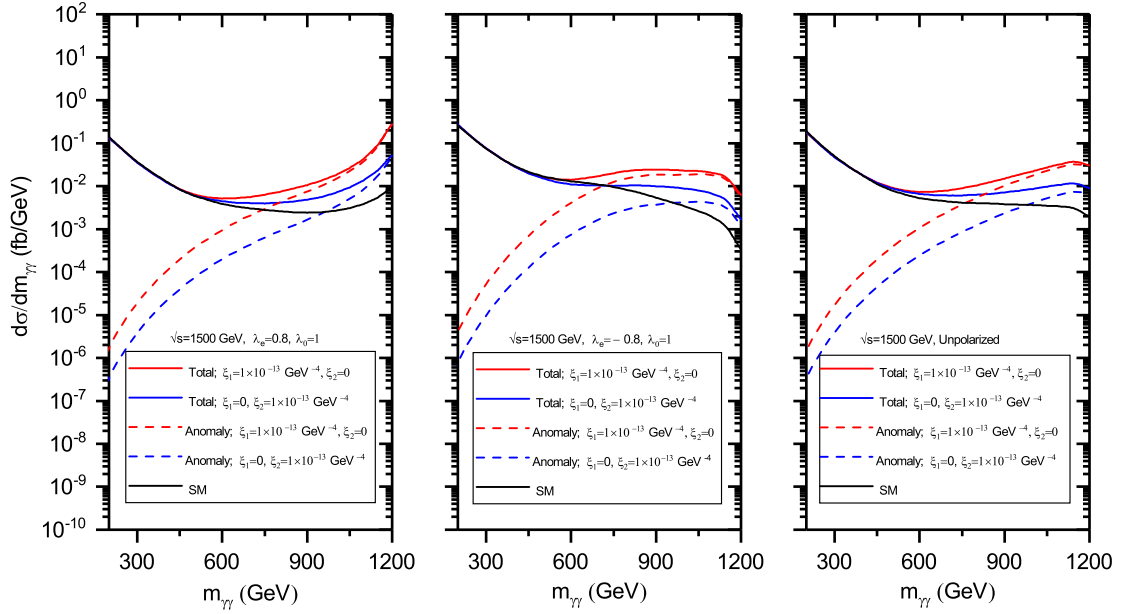}
\caption{The differential cross sections for the process
$\gamma\gamma\rightarrow\gamma\gamma$ as functions of the invariant
mass of the outgoing photons for the $e^+e^-$ collider energy
$\sqrt{s} = 1500$ GeV. The left, middle and right panels correspond
to the electron beam helicitiy $\lambda_e = 0.8, -0.8$, and 0,
respectively. The curves on each plot (from the top downwards) are:
the differential cross sections for the coupling sets ($\zeta_1 =
10^{-13} \mathrm{\ GeV}^{-4}$, $\zeta_2 = 0$) and ($\zeta_1 = 0$,
$\zeta_2 = 10^{-13} \mathrm{\ GeV}^{-4}$), the anomalous
contribution for the same coupling values, the SM cross section.}
\label{fig:WDE750}
\end{center}
\end{figure}

\begin{figure}[htb]
\begin{center}
\includegraphics[scale=0.6]{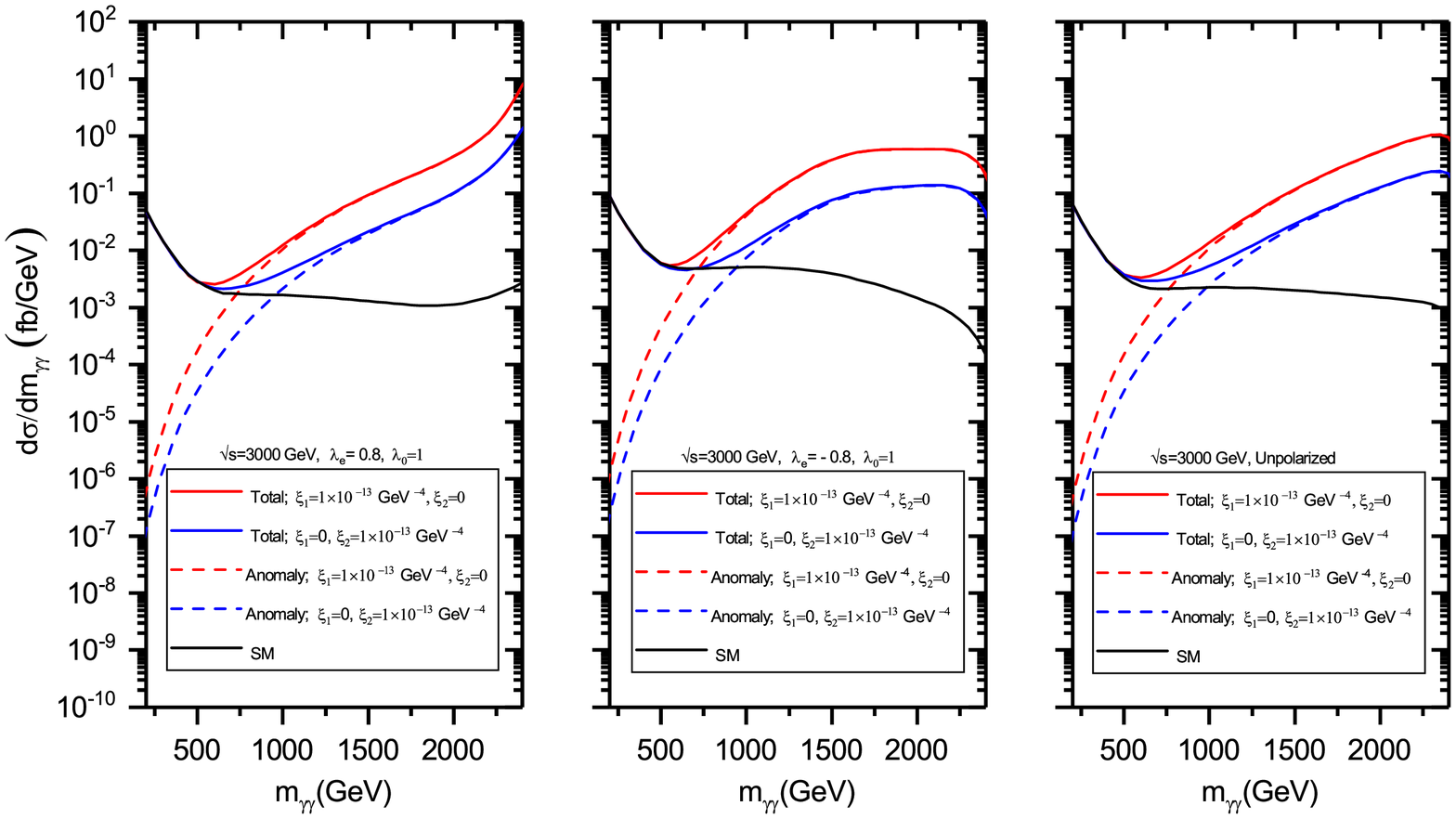}
\caption{The same as in Fig.~\ref{fig:WDE750}, but for the $e^+e^-$
collider energy $\sqrt{s} = 3000$ GeV.}
\label{fig:WDE1500}
\end{center}
\end{figure}

The differential cross sections as functions of the photon invariant
mass $m_{\gamma\gamma}$ are shown in Figs.~\ref{fig:WDE750} and
\ref{fig:WDE1500}.
%
%
We imposed the cut on the rapidity of the outgoing photons,
$|\eta_{\gamma\gamma}| < 2.5$. The left, middle and rights panels of
these figures correspond to the electron beam helicities $\lambda_e
= 0.8$, $\lambda_e = -0.8$, and $\lambda_e = 0$, respectively. Note
that the anomaly amplitude is pure real, while the SM one is mainly
imaginary. As a result, the interference contribution to the
differential cross section is relatively very small for any values
of $m_{\gamma\gamma}$ in the region $m_{\gamma\gamma} > 200$ GeV.
If, for instance, $\sqrt{s} = 1500$ GeV, $\lambda_e = 0.8$, $\zeta_1
= 10^{-13} \mathrm{\ GeV}^{-4}$, $\zeta_2 = 0$, and
$m_{\gamma\gamma} = 500$ GeV, the anomaly, SM, and interference
terms of the cross section are equal to $3.45 \times 10^{-4}$
fb/GeV, $5.90 \times 10^{-3}$ fb/GeV, and $9.05 \times 10^{-5}$
fb/GeV, respectively. For $\sqrt{s} = 3000$ GeV, the same values of
$\lambda_e$, $\zeta_{1,2}$, and $m_{\gamma\gamma} = 1000$ GeV we
find, correspondingly, $1.05 \times 10^{-2}$ fb/GeV, $1.64 \times
10^{-3}$ fb/GeV, and $1.80 \times 10^{-4}$ fb/GeV.

For both $\sqrt{s}$, and any value of $\lambda_e$, the anomaly
differential cross sections become to dominate the SM background at
about $m_{\gamma\gamma} > 750$ GeV for $\zeta_1 = 10^{-13} \mathrm{\
GeV}^{-4}$, $\zeta_2 = 0$. For the couplings $\zeta_1 = 0$, $\zeta_2
= 10^{-13} \mathrm{\ GeV}^{-4}$ it takes place in the region
$m_{\gamma\gamma} > 960$ GeV. For the same $\sqrt{s}$ and
$\zeta_{1,2}$, the differential cross section with $\lambda_e = 0.8$
becomes larger than the differential cross section with the opposite
beam helicity $\lambda_e = -0.8$ and unpolarized one, as
$m_{\gamma\gamma}$ grows. A possible background with fake photons
from decays of $\pi^0$, $\eta$, and $\eta'$ is negligible in the
signal region.

The leading part of the anomalous cross section is proportional to
$s^2$. However, it does not mean that the unitarity is violated for
the region of the anomalous QGCs considered in our paper. As it is
shown in \cite{Almeida:2020}, the anomalous quartic couplings of the
order of $10^{-13}$ GeV$^{-4}$ do not lead to unitarity violation
for the collision energy below 3 TeV.

\begin{figure}[htb]
\begin{center}
\includegraphics[scale=0.6]{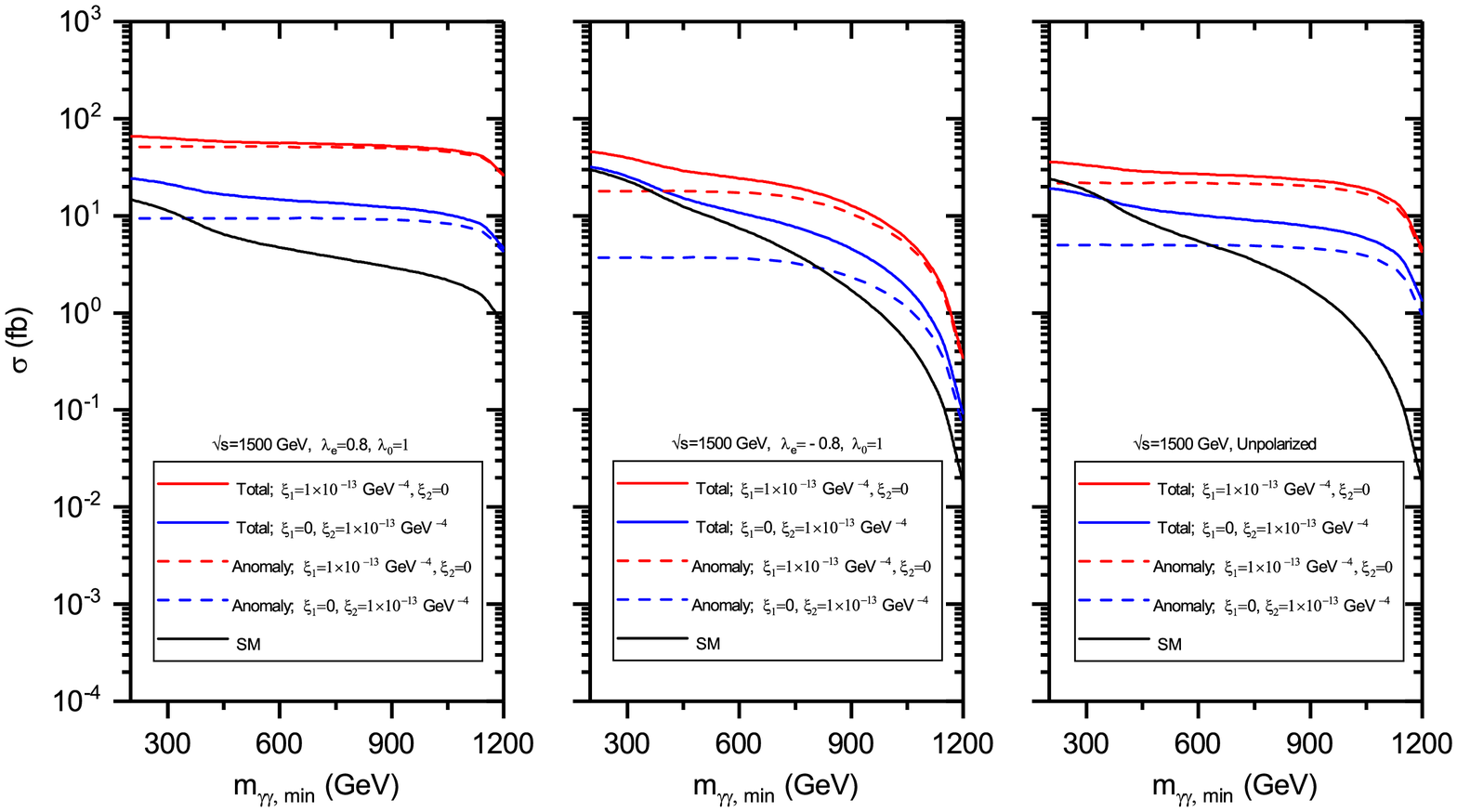}
\caption{The total cross sections for the process
$\gamma\gamma\rightarrow\gamma\gamma$ as functions of the minimal
invariant mass of the outgoing photons for the $e^+e^-$ collider
energy $\sqrt{s} = 1500$ GeV. The left, middle and right panels
correspond to the electron beam helicitiy $\lambda_e = 0.8, -0.8$,
and 0, respectively. The curves on each plot (from the top
downwards) are: the total cross sections for the coupling sets
($\zeta_1 = 10^{-13} \mathrm{\ GeV}^{-4}$ , $\zeta_2 = 0$) and
($\zeta_1 = 0$, $\zeta_2 = 10^{-13} \mathrm{\ GeV}^{-4}$), the
anomalous contribution for the same coupling values, the SM cross
section.}
\label{fig:WCUTE750}
\end{center}
\end{figure}
%
\begin{figure}[htb]
\begin{center}
\includegraphics[scale=0.6]{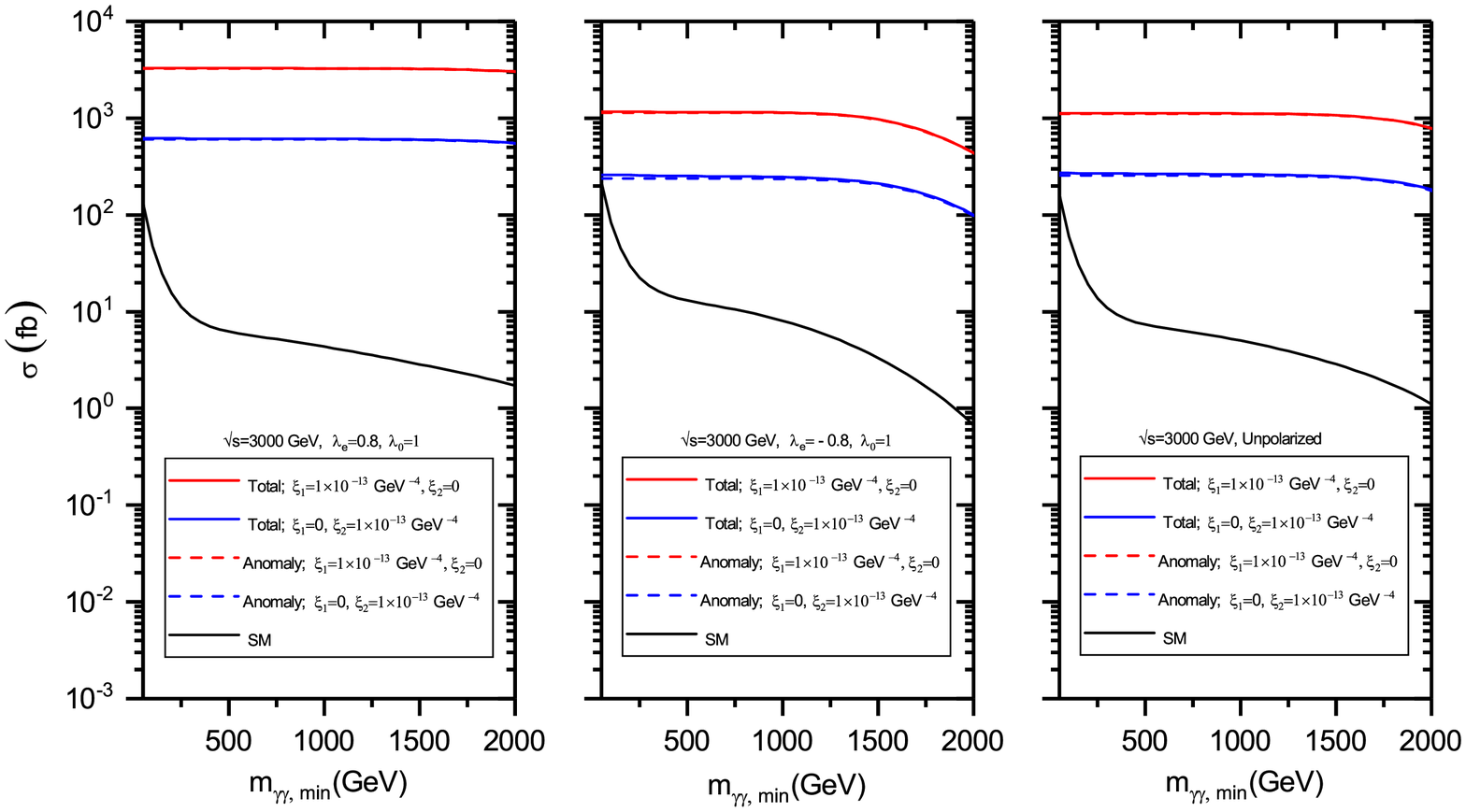}
\caption{The same as in Fig.~\ref{fig:WCUTE750}, but for the
$e^+e^-$ collider energy $\sqrt{s} = 3000$ GeV.}
\label{fig:WCUTE1500}
\end{center}
\end{figure}

The results of our calculations of the total cross sections
$\sigma(m_{\gamma\gamma} > m_{\gamma\gamma,\min})$, where
$m_{\gamma\gamma,\min}$ is the minimal invariant mass of the
outgoing photons, are shown in
Figs.~\ref{fig:WCUTE750}-\ref{fig:WCUTE1500}. The results are
presented for two values of the CLIC energy, and two sets of the
couplings $\zeta_1, \zeta_2$. The reader can obtain the prediction
for any value of the coefficients $\zeta_1, \zeta_2$ by simply
rescaling the results. For $\sqrt{s} = 1500$ GeV, $\lambda_e = 0.8$,
$\zeta_1 = 10^{-13}$ GeV$^{-4}$, and $\zeta_2 = 0$, the total cross
section remains almost unchanged despite increasing
$m_{\gamma\gamma,\min}$. A similar tendency takes place for the
unpolarized cross section. On the contrary, for $\lambda_e = -0.8$,
the total cross sections decrease rapidly, as
$m_{\gamma\gamma,\min}$ grows. For $\sqrt{s} = 3000$ GeV the total
cross section deviation from the SM gets higher, as
$m_{\gamma\gamma, \min}$ increases. $\sigma(m_{\gamma\gamma} >
m_{\gamma\gamma,\min})$ with $\lambda_e = 0.8$ and the unpolarized
cross section are almost independent of $m_{\gamma\gamma, \min}$,
while $\sigma(m_{\gamma\gamma} > m_{\gamma\gamma,\min})$ with
$\lambda_e = -0.8$ decreases at large $m_{\gamma\gamma, \min}$. The
total cross section with $\lambda_e = 0.8$ is several times large
than the total cross section with the opposite beam helicity. Note,
however, that for $\lambda_e = 0.8$ the CLIC expected integrated
luminosities are four times smaller than those for $\lambda_e =
-0.8$, for both values of $e^+e^-$ collision energy, see
Tab.~\ref{tab:1}.

To calculate the exclusion region, we use the following formula for
the exclusion significance \cite{Zhang:2020}
\begin{equation}\label{S_excl}
S_{\mathrm{excl}} = \sqrt{ 2\left[ (s - b \ln \left( \frac{b + s +
x}{2b} \right) - \frac{1}{\delta^2}\ln \left( \frac{b - s + x}{2b}
\right) - (b + s -x) \left( 1 + \frac{1}{\delta^2 b} \right) \right]
} \;,
\end{equation}
with
\begin{equation}\label{x}
x = \sqrt{(s+b)^2 - 4\delta^2 s b^2/(1 + \delta^2 b)} \;.
\end{equation}
Here $s$ and $b$ represent the total number of signal and background
events, respectively, and $\delta$ is the percentage systematic
error. In the limit $\delta \rightarrow 0$ expression \eqref{S_excl}
is simplified to be
\begin{equation}\label{S_excl_no_sist}
S_{\mathrm{excl}} = \sqrt{2[s - b\,\ln(1 + s/b)]} \;.
\end{equation}
We define the regions $S_{\mathrm{excl}} \leqslant 1.645$ as the
regions that can be excluded at the 95\% C.L.

\begin{figure}[htb]
\begin{center}
\includegraphics[scale=0.6]{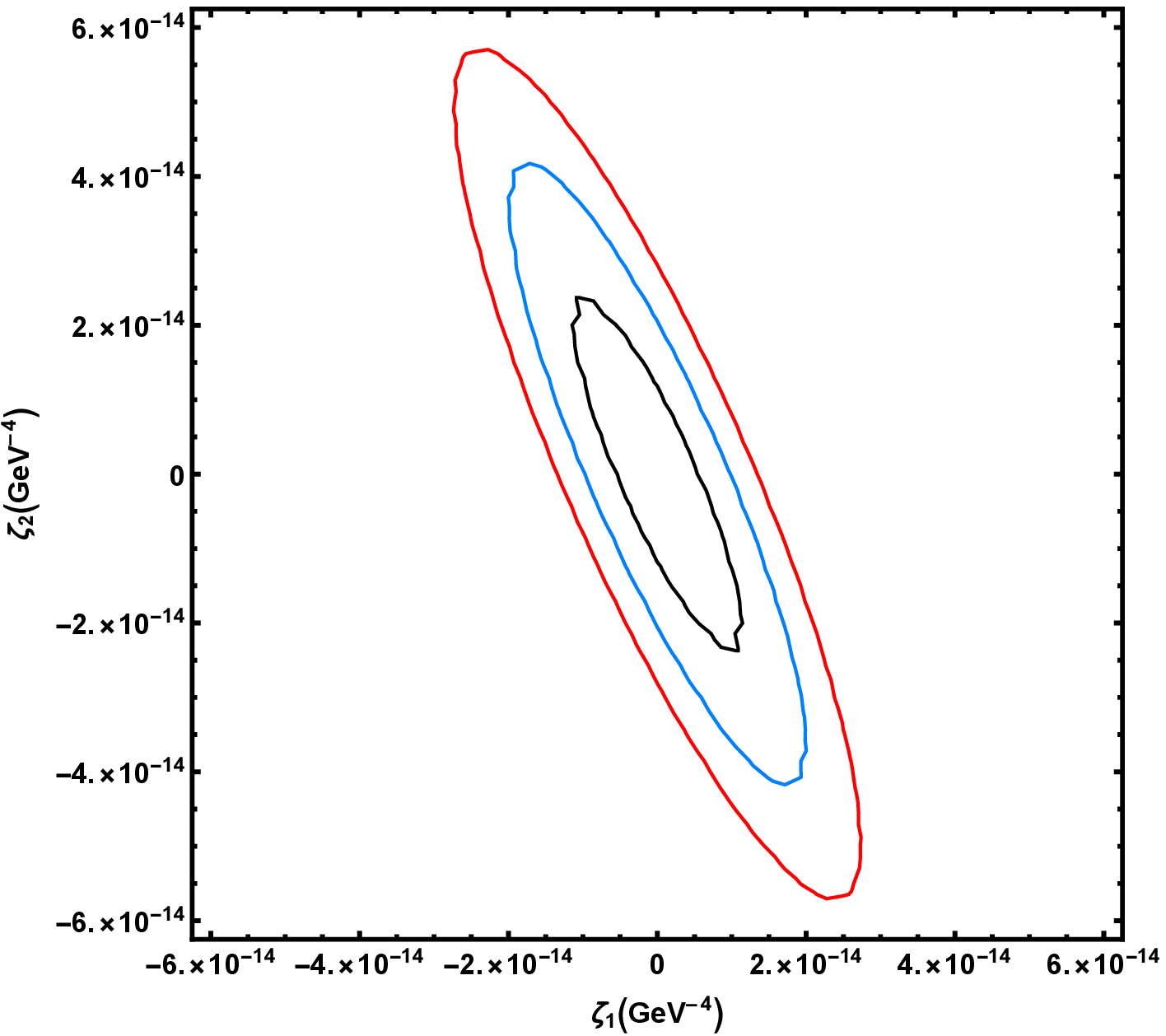}
\caption{The 95\% C.L. exclusion regions for the couplings $\zeta_1,
\zeta_2$ for the unpolarized light-by-light scattering at the CLIC
with the systematic errors $\delta = 0\%$ (black ellipse), $\delta =
5\%$ (blue ellipse), and $\delta = 10\%$ (red ellipse). The inner
regions of the ellipses are inaccessible. The collision energy is
$\sqrt{s} = 1500$ GeV, the integrated luminosity is $L = 2500$
fb$^{-1}$. The cut on the outgoing photon invariant mass
$m_{\gamma\gamma} > 1000$ GeV was imposed.}
\label{fig:SSE750}
\end{center}
\end{figure}

\begin{figure}[htb]
\begin{center}
\includegraphics[scale=0.6]{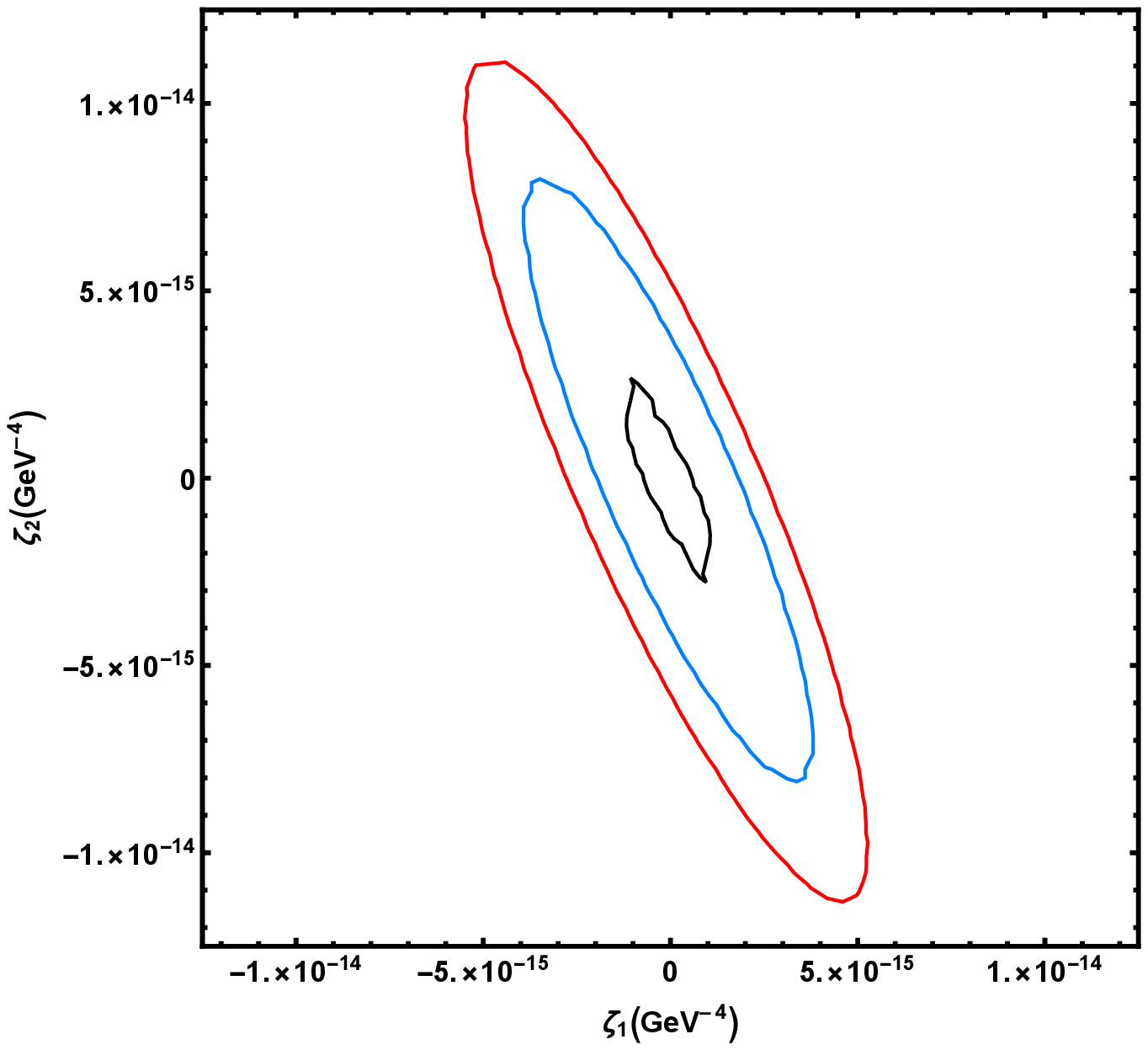}
\caption{The same as in Fig.~\ref{fig:SSE750}, but for $\sqrt{s} =
3000$ GeV and $L=5000$ fb$^{-1}$.}
\label{fig:SSE1500}
\end{center}
\end{figure}

Our 95\% C.L. exclusion regions for the couplings $\zeta_1, \zeta_2$
for the unpolarized LBL scattering are shown in
Figs.~\ref{fig:SSE750}, \ref{fig:SSE1500} with the cuts
$|\eta_{\gamma\gamma}| < 2.5$, $m_{\gamma\gamma} > 1000$ GeV, for
$\delta = 0$, $\delta = 5\%$, and $\delta = 10\%$. Note that for the
unpolarized process the pure anomaly cross section is proportional
to the coupling combination $48\zeta_1^2 + 40\zeta_1 \zeta_2 +
11\zeta_2^2$ \cite{Fichet:2015}. As a result, the exclusion regions
are ellipses rotated counterclockwise in the plane $(\zeta_1,
\zeta_2)$ through the angle $0.5\arctan(80/37) \simeq 32.6^\circ$
about the origin.

%

For the polarized LBL scattering at the CLIC, the exclusion bounds
on the anomalous photon couplings are presented in
Tabs.~\ref{tab:2}, \ref{tab:3} using the cut $m_{\gamma\gamma} >
1000$ GeV. Note that the values of the expected integrated
luminosities depend on the energy $\sqrt{s}$. As one can see from
these tables, for both energies the exclusion bounds on couplings
$\zeta_1$ and $\zeta_2$ weakly depend on the helicity of the initial
electron beams.
\begin{table}[h]
\centering
\begin{tabular}{||l||c|c|c|c||}
  \hline
  Helicity & & 0 & $-0.8$ & 0.8 \\
  \hline
  Luminosity, fb$^{-1}$ & & 2500 & 2000 & 500 \\
  \hline
  $|\zeta_1|, \mathrm{GeV}^{-4} \ (\zeta_2=0)$ & \makecell{$\delta=0\%$ \\  $\delta=5\%$ \\ $\ \, \delta=10\%$ } &
  $\makecell{ 8.52\times10^{-15} \\ 9.84\times10^{-15} \\1.36\times10^{-14}}$ &
  $\makecell{ 7.35\times10^{-15} \\ 1.06\times10^{-14} \\1.42\times10^{-14}}$  & $\makecell{ 7.32\times10^{-15} \\ 9.59\times10^{-15} \\1.26\times10^{-14}}$ \\
  \hline
  $|\zeta_2|, \mathrm{GeV}^{-4} \ (\zeta_1=0)$& \makecell{$\delta=0\%$ \\  $\delta=5\%$ \\ $\ \, \delta=10\%$ } &
  $\makecell{ 1.71\times10^{-14} \\ 2.06\times10^{-14} \\2.81\times10^{-14}}$ & $\makecell{ 1.45\times10^{-14} \\ 2.19\times10^{-14} \\2.95\times10^{-14}}$ &
  $\makecell{ 1.24\times10^{-14} \\ 2.01\times10^{-14} \\2.60\times10^{-14}}$ \\
  \hline
\end{tabular}
\caption{The 95\% C.L. exclusion limits on the couplings $\zeta_1$
and $\zeta_2$ for the CLIC collision energy $\sqrt{s} = 1500$ GeV,
and the cut $m_{\gamma\gamma} > 1000$ GeV.}
\label{tab:2}
\end{table}
\begin{table}[h]
\centering
\begin{tabular}{||l||c|c|c|c||}
  \hline
  Helicity & & 0 & $-0.8$ & 0.8 \\
  \hline
  Luminosity, fb$^{-1}$ & & 5000 & 4000 & 1000 \\
  \hline
  $|\zeta_1|, \mathrm{GeV}^{-4} \ (\zeta_2=0)$ & \makecell{$\delta=0\%$ \\ $\delta=5\%$ \\ $\ \, \delta=10\%$ } &
  $\makecell{ 6.85\times10^{-16} \\ 1.90\times10^{-15} \\2.63\times10^{-15}}$ &
  $\makecell{ 8.82\times10^{-16} \\ 2.48\times10^{-15} \\3.37\times10^{-15}}$ & $\makecell{ 8.73\times10^{-16} \\ 1.56\times10^{-15} \\2.12\times10^{-15}}$ \\
  \hline
  $|\zeta_2|, \mathrm{GeV}^{-4} \ (\zeta_1=0)$& \makecell{$\delta=0\%$ \\  $\delta=5\%$ \\ $\ \, \delta=10\%$ } &
  $\makecell{ 1.43\times10^{-15} \\ 3.99\times10^{-15} \\5.53\times10^{-15}}$ & $\makecell{ 1.85\times10^{-15} \\ 5.12\times10^{-15} \\7.10\times10^{-15}}$ &
  $\makecell{ 1.82\times10^{-15} \\ 3.28\times10^{-15} \\4.46\times10^{-15}}$ \\
  \hline
\end{tabular}
\caption{The same as in Tab.~\ref{tab:2}, but for $\sqrt{s} = 3000$
GeV.}
\label{tab:3}
\end{table}


\begin{figure}[htb]
\begin{center}
\includegraphics[scale=0.6]{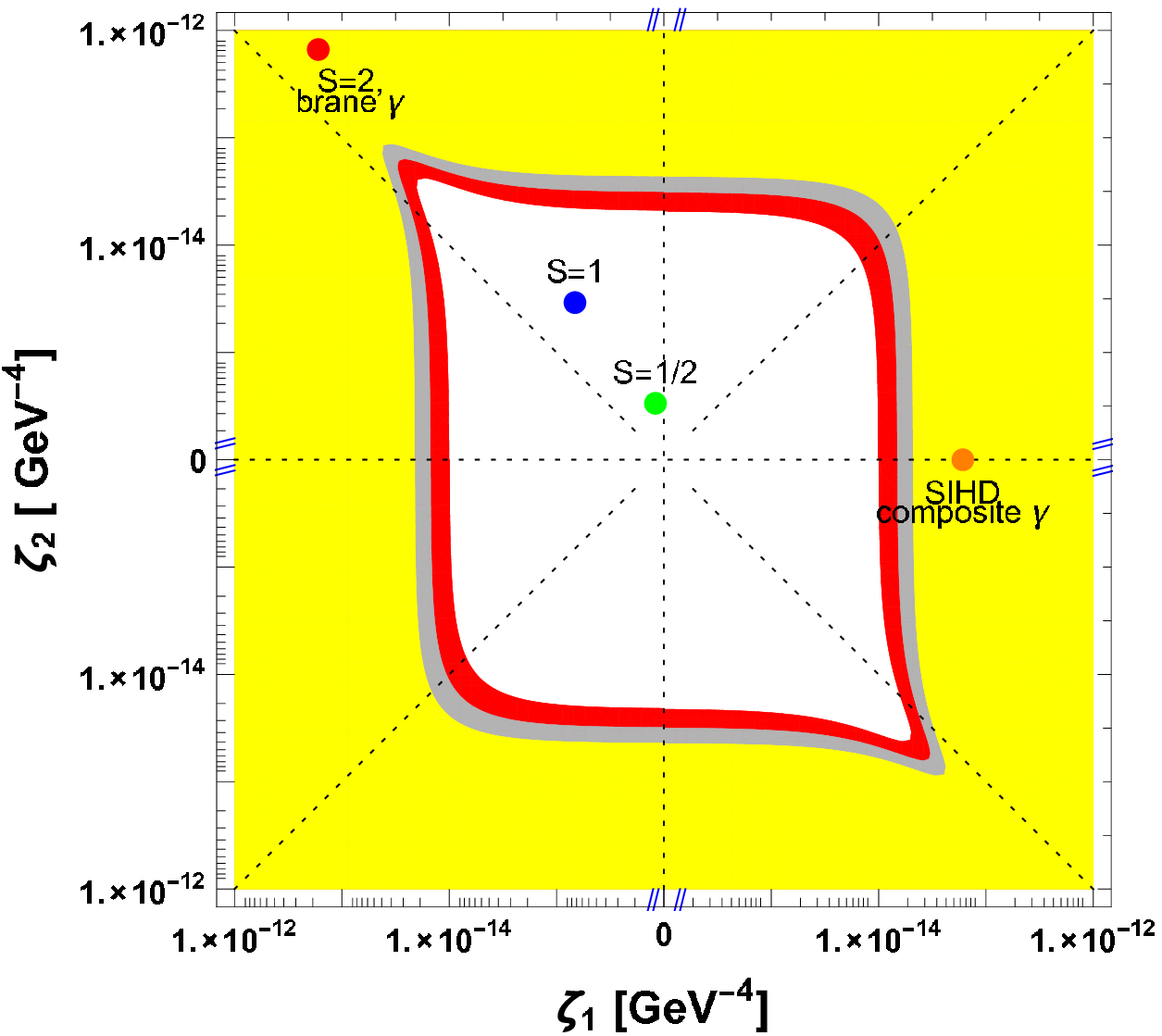}
\caption{The LHC sensitivity in the $(\zeta_1, \zeta_2)$ plane. In
particular, the red region can be probed at the 95\% C.L. using
proton tagging at the LHC. The white region is inaccessible. The
figure is taken from Ref.~\protect\cite{Fichet:2015}.}
\label{fig:zeta_plot}
\end{center}
\end{figure}

Previously, the discovery potential for the LBL scattering at the 14
TeV LHC has been estimated in
\cite{Enterria:2013}-\cite{Fichet:2015}. As was shown in
\cite{Fichet:2014_2}, the 14 TeV LHC 95\% C.L. exclusion limits on
$\zeta_1$ and $\zeta_2$ couplings are $1.5 \times 10^{-14}$
GeV$^{-4}$ and  $3.0 \times 10^{-14}$ GeV$^{-4}$, respectively, for
$L = 300$ fb$^{-1}$ integrated luminosity. For $L = 3000$ fb$^{-1}$
(HL-LHC), the values are twice smaller, $7.0 \times 10^{-15}$
GeV$^{-4}$ and $1.5 \times 10^{-14}$ GeV$^{-4}$. The sensitivity in
the $(\zeta_1, \zeta_2)$ plane is shown in Fig.~\ref{fig:zeta_plot}
taken from \cite{Fichet:2015}. As one can see from Tab.~\ref{tab:2},
our CLIC bounds on the couplings $\zeta_1, \zeta_2$ for the LBL
scattering with $\sqrt{s} = 1500$ GeV are comparable with the HL-LHC
bounds~\cite{Fichet:2015}. However, for $\sqrt{s} = 3000$ GeV our
lower bounds on $\zeta_1, \zeta_2$ are approximately one order of
magnitude smaller than the HL-LHC ones, see Tab.~\ref{tab:3}.

In \cite{Koksal:2016} the CLIC 95\% C.L. sensitivity bounds on the
coefficients $f_{T_0}/\Lambda^4$ and $f_{T_9}/\Lambda^4$ in the EFT
Lagrangian (equivalent to the coefficients $c_9/\Lambda^4$ and
$c_{13}/\Lambda^4$ in \eqref{Lagrangian_QNGCs}) for $\sqrt{s} =
3000$ GeV and $L = 2000$ fb$^{-1}$ are presented. Note that these
coefficients are only parts of our couplings $\zeta_1$ and $\zeta_2$
\eqref{zeta_c}. The bounds have been obtained by examining the
anomalous quartic couplings of $ZZ\gamma\gamma$ vertex.

\section{Conclusions} %

In the present paper, we have examined the anomalous quartic neutral
couplings of the $\gamma\gamma\gamma\gamma$ vertex in the polarized
light-by-light collisions of the Compton backscattered photons at
the CLIC. Both the second and third stages of the CLIC are
considered with the collision energies $\sqrt{s} = 1500$ GeV and
$\sqrt{s} = 3000$ GeV, respectively. The helicity of the initial
electron beam was taken to be $\lambda_e = \pm \,0.8$. The
unpolarized case ($\lambda_e = 0$) has been also considered. We used
the $SU(2)_L \times U(1)_Y$ effective Lagrangian describing the
contribution to the anomalous quartic neutral gauge boson couplings.
Its part, relevant to the anomalous $\gamma\gamma\gamma\gamma$
vertex \eqref{contact_Lagrangian}, expressed in terms of the
physical fields, contains two couplings $\zeta_1$, $\zeta_2$ of
dimension $-4$.

We have calculated both the differential and total cross sections of
the light-by-light scattering $\gamma\gamma\rightarrow\gamma\gamma$,
with the cut imposed on the rapidity of the final photons,
$|\eta_{\gamma\gamma}| < 2.5$. The plots for two values of the
collision energy $\sqrt{s}$ and three values of the electron beam
helicity $\lambda_e$ (including the unpolarized case) are presented.
The anomaly and SM contributions to the cross sections are presented
separately. The CLIC exclusion sensitivity bounds on the anomaly
coupling constants $\zeta_1$ and $\zeta_2$, coming from the process
$\gamma\gamma\rightarrow\gamma\gamma$, are calculated for three
values of the systematic error, $\delta = 0\%$, $\delta = 5\%$, and
$\delta = 10\%$. To reduce the SM background, we imposed the cut on
the invariant mass of the outgoing photons, $m_{\gamma\gamma} >
1000$ GeV.

For the unpolarized LBL scattering at the CLIC, the 95\% C.L.
exclusion regions are shown in Figs.~\ref{fig:SSE750},
\ref{fig:SSE1500}. The exclusion bounds for the polarized LBL
scattering are presented in Tabs.~\ref{tab:2}, \ref{tab:3} for three
values of the systematic error. For the $e^+e^-$ collision energy
$\sqrt{s} = 3000$ GeV, electron beam helicity $\lambda = 0.8$, and
$\delta = 10\%$, our bounds on $\zeta_1, \zeta_2$ have appeared to
be approximately one order of magnitude stronger than the
corresponding HL-LHC bounds obtained for $\sqrt{s} = 14$ TeV and
integrated luminosity $L = 3000$ fb$^{-1}$ in \cite{Fichet:2015}.

All said above allows us to conclude that the LBL scattering at the
CLIC, especially the polarized, has a great physical potential in
searching for the anomalous quartic neutral couplings of the
$\gamma\gamma\gamma\gamma$ vertex.





\begin{thebibliography}{99}
%
\bibitem{Buchmuller:1986}
W.~Buchmuller and D.~Wyler, \emph{Effective lagrangian analysis of
new interactions and flavour conservation}, Nucl. Phys. B
\textbf{268}, 621 (1986).
\bibitem{Higawara:1987}
K.~Hagiwara, R.D.~Peccei, D.~Zeppenfeld and K.~Hikasa, \emph{Probing
the weak boson sector in $e^+e^- \rightarrow W^+W^-$}, Nucl. Phys. B
\textbf{282}, 253 (1987).
%
\bibitem{Godfrey:1995}
S.~Godfrey, \emph{Quartic gauge boson couplings}, in Proceedings of
the International Simposium on Vector Boson Self-Interactions, Los
Angeles, CA, USA, 1-3 February 1995, pp.~209-223
(arXiv:hep-ph/9505252).
%
\bibitem{Balange:1992}
G.~Balanger and F.~Boudjema, \emph{$\gamma\gamma\rightarrow W^+W^-$
and $\gamma\gamma\rightarrow ZZ$ as tests of novel quartic
couplings}, Phys. Lett. B. \textbf{288}, 210 (1992).
%
\bibitem{Stirling:1999}
W.J.~Stirling and A.~Werhenbach, \emph{Anomalous quartic couplings
in $\nu\bar{\nu}\gamma\gamma$ production via $WW$-fusion at LEP2},
Phys. Lett. B \textbf{466}, 369 (1999); W.J.~Stirling and
A.~Werhenbach, \emph{Anomalous quartic couplings in $W^+W^-\gamma$,
$Z^0Z^0\gamma$ and $Z^0\gamma\gamma$ production at present and
future $e^+e^-$ colliders}, Eur. Phys. J. C \textbf{14}, 103 (2000).
%
\bibitem{Rujula:1992}
A.~De~R\'{u}jula, M.B.~Gavela, P.~Hernandez and E.~Mass\'{o},
\emph{The self-couplings of vector bosons: does LEP-1 obviate
LET-2?}, Nucl. Phys. B \textbf{384}, 3 (1992).
%
\bibitem{Hagiwara:1992}
K.~Hagiwara, S.~Ishihara, R.~Szalapski and D.~Zeppenfeld,
\emph{Low-energy constraints on electroweak three gauge boson
couplings}, Phys. Lett. B \textbf{283}, 353 (1992); \emph{Low energy
effects of new interactions in the electroweak boson sector}, Phys.
Rev. D \textbf{48}, 2182 (1993).
%
\bibitem{Degrande:2013}
C.~Degrande \emph{et al.}, \emph{Effective field theory: a modern
approach to anomalous couplings}, Ann. Phys. \textbf{335}, 21
(2013).
\bibitem{Eboli:2006}
O.J.P.~\'{E}boli, M.C.~Gonzalez-Garcia and J.K.~Mizukoshi, \emph{$pp
\rightarrow jje^\pm\mu^\pm \nu\nu$ and $jje^\pm\mu^\mp \nu\nu$ at
$\mathcal{O}(\alpha^6_{\mathrm{em}})$ and
$\mathcal{O}(\alpha^4_{\mathrm{em}} \alpha_s^2)$ for the study of
the quartic electroweak gauge boson vertex at CERN LHC}, Phys. Rev.
D \textbf{74}, 073005 (2006).
%
\bibitem{Gupta:2012}
R.S.~Gupta, \emph{Probing quartic gauge boson couplings using
diffractive photon fusion at the LHC}, Phys. Rev. D \textbf{85},
014006 (2012).
%
\bibitem{Fichet:2014_1}
S.~Fichet and G.~von~Gersdorff, \emph{Anomolous gauge couplings from
composite Higgs and warped extra dimensions}, JHEP \textbf{03}, 102
(2014).
%
\bibitem{LEP_1}
P.~Achand \emph{et al.} (L3 Collaboration), \emph{Study of the
$W^+W^-\gamma$ process and limits on anomalous quartic gauge boson
couplings at LEP}, Phys. Lett. B \textbf{527}, 29 (2002); A.~Heister
\emph{et al.} (ALEPH Collaboration), \emph{Constraints on anomalous
QGC's in $e^+e^-$ interactions from 183 GeV to 209 GeV}, Phys. Lett.
B \textbf{602}, 31 (2004); G.~Abbiendi \emph{et al.} (OPAL
Collaboration), \emph{A study of $W^+W^-\gamma$ events at LEP},
Phys. Lett. B \textbf{580}, 17 (2004); J.~Abdallah \emph{et al.}
(DELPHI Collaboration), \emph{Measurement of the $e^+e^- \rightarrow
W^+W^-\gamma$ cross-section and limits on anomalous quartic gauge
couplings with DELPHI}, Eur. Phys. J. C \textbf{31}, 139 (2003).
%
\bibitem{Belanger:2000}
G.~B\'{e}langer \emph{et al.}, \emph{Bosonic quartic couplings at
LEP-2}, Eur. Phys. J. C \textbf{13}, 283 (2000).
%
\bibitem{Tevatron}
V.M.~Abazov \emph{et al.} (D0 Collaboration), \emph{Search for
anomalous quartic $WW\gamma\gamma$ couplings in dielectron and
missing energy final states in $\bar{p}p$ collisions at
$\sqrt{s}=1.96$ TeV}, Phys. Rev. D \textbf{88}, 012005 (2013).
%
\bibitem{LEP_2}
P.~Achand \emph{et al.} (L3 Collaboration), \emph{The $e^+e^-
\rightarrow Z \gamma\gamma \rightarrow q \bar{q} \gamma\gamma$
reaction at LEP and constraints on anomalous quartic gauge boson
couplings}, Phys. Lett. B \textbf{540}, 43 (2002).
%
\bibitem{Eboli:2001_1}
O.J.P.~\'{E}boli, M.C.~Gonzalez-Garcia, S.M.~Lietti and S.F.~Novaes,
\emph{Anomalous quartic gauge boson couplings at hadron colliders},
Phys. Rev. D \textbf{63}, 075008 (2001); O.J.P.~\'{E}boli,
M.C.~Gonzalez-Garcia and S.M.~Lietti, \emph{Bosonic quartic
couplings at CERN LHC}, Phys. Rev. D \textbf{69}, 095005 (2004).
%
\bibitem{Pierzchala:2008}
T.~Pierzcha{\l}a and K.~Piotrzkowski, \emph{Sensitivity to anomalous
quartic gauge couplings in photon-photon interactions at the LHC},
Nucl. Phys. B. Proc. Suppl. \textbf{179}, 257 (2008).
%
\bibitem{Chapon:2009}
E.~Chapon, O.~Kepka and C.~Royon, \emph{Probing $WW\gamma\gamma$ and
$ZZ\gamma\gamma$ quartic anomalous couplings with 10 pb$^{-1}$ at
the LHC}, arXiv:0908.1061; E.~Chapon, C.~Royon and O.~Kepka,
\emph{Anomalous quartic $WW\gamma\gamma$, $ZZ\gamma\gamma$, and
trilinear $WW\gamma$ couplings in two-photon processes at high
luminosity at the LHC}, Phys. Rev. D \textbf{81}, 074003 (2010).
%
\bibitem{Senol:2014}
A.~Senol, \emph{Anomalous quartic $WW\gamma\gamma$ and
$ZZ\gamma\gamma$ couplings in $\gamma p$ collision at the LHC}, Int.
J. Mod. Phys. A \textbf{29}, 1450148 (2014).
%
\bibitem{Perez:2018}
G.~Perez, M.~Sekulla and D.~Zeppenfeld, \emph{Anomalous quartic
gauge couplings and unitarization for the vector boson scattering
process $pp \rightarrow W^+W^+ jj X \rightarrow l^+\nu_l l^+\nu_l
jjX$}, Eur. Phys. J. C \textbf{78}, 759 (2018).
%
\bibitem{Guo:2020}
Yu-Chen~Guo, Ying-Ying~Wang, Ji-Chong~Yang and Chong-Xing~Yue,
\emph{Constraints on anomalous quartic gauge couplings via $W\gamma
jj$ production at the LHC}, Chin. Phys. C \textbf{44}, 123105
(2020).
%
\bibitem{Tizchang:2020}
S.~Tizchang and S.~M.~Etesami, \emph{Pinning down the gauge boson
couplings in $WW\gamma$ production using forward proton tagging},
JHEP \textbf{07}, 191 (2020).
%
\bibitem{Zhu:2020}
Jian-Wen~Zhu, Ren-You~Zhang, Wen-Gan~Ma, Q.~Yang and Yi~Jiang,
\emph{$W^+W^-\gamma$ production at hadron colliders with NLO QCD+EW
corrections and parton shower effects}, J. Phys. G: Nucl. Part.
Phys. \textbf{47}, 055006 (2020).
%
%
\bibitem{Sahin:2012}
\.{I}.~\c{S}ahin and B.~\c{S}ahin, \emph{Anomalous quartic
$ZZ\gamma\gamma$ couplings in gamma-proton collision at the LHC},
Phys. Rev. D \textbf{86}, 115001 (2012).
%
\bibitem{CMS:QGCs}
A.M.~Sirunyuan \emph{et al.} (CMS Collaboration), \emph{Measurement
of the cross section for electroweak production of a $Z$ boson, a
photon and two jets in proton-proton collisions at $\sqrt{s} = 13$
TeV and constraints on anomalous quartic couplins}, JHEP
\textbf{06}, 076 (2020).
\bibitem{ATLAS:QGCs}
G.~Aad \emph{et al.} (ATLAS Collaboration), \emph{Measurements of
$Z\gamma$ and $Z\gamma\gamma$ production in $pp$ collisions at
$\sqrt{s} = 8$ TeV with the ATLAS detector}, Phys. Rev. D
\textbf{93}, 112002 (2006).
%
\bibitem{Ari:2020_1}
V.~Ari, A.~Guti\'{e}rrez-Rodriguez, M.A.~Hernandez-Ruiz and
M.~K\"{o}ksal, \emph{Anomalous quartic $W^+W^-\gamma\gamma$
couplings in $e^-p$ collisions at the LHeC and the FCC-he}, Eur.
Phys. J. Plus \textbf{135}, Article number: 336 (2020).
%
\bibitem{Ari:2020_2}
V.~Ari, E.~Gurkanli, A.A.~Billur and M.~K\"{o}ksal, \emph{Model
independent study for the anomalous quartic $WW\gamma\gamma$
couplings at future electron-proton colliders}, Nucl. Phys. B
\textbf{957}, 115102 (2020).
%
\bibitem{Rodriguez:2021}
A.~Gutierrez-Rodriguez, M.A.~Hernandez-Ruiz, E.~Gurkanli, V.~Ari and
M.~K\"{o}ksal, \emph{Study on the anomalous quartic
$W^+W^-\gamma\gamma$ couplings of electroweak bosons in $e^-p$
collisions at the LHeC and the FCC-he}, Eur. Phys. J. C \textbf{81},
210 (2021).
%
\bibitem{Denner:2001}
A.~Denner, S.~Dittmaier, M.~Roth and D.~Wackeroth, \emph{Probing
anomalous quartic gauge-boson couplings via $e^+e^- \rightarrow 4$
fermions + $\gamma$}, Eur. Phys. J. C \textbf{20}, 201 (2001).
%
\bibitem{Eboli:1994}
O.J.P.~\'{E}boli, M.C.~Gonzalez-Garcia and S.F.~Novaes,\emph{Quartic
anomalous couplings in $e\gamma$ colliders}, Nucl. Phys. B
\textbf{411}, 381 (1994).
%
\bibitem{Atag:2007}
S.~Ata\u{g} and \.{I}.~\c{S}ahin, {Anomalous quartic
$WW\gamma\gamma$ and $ZZ\gamma\gamma$ couplings in $e\gamma$
collision with initial beams and final state polarizations}, Phys.
Rev. D \textbf{75}, 073003 (2007).
%
\bibitem{Eboli:2001_2}
O.J.P.~\'{E}boli and J. K. Mizukoshi, \emph{Probing anomalous
quartic couplings in $e\gamma$ and $\gamma\gamma$ colliders}, Phys.
Rev. D \textbf{64}, 075011 (2001).
\bibitem{Sahin:2009}
\.{I}.~\c{S}ahin, \emph{Anomalous quartic $WW\gamma\gamma$ and
$WWZ\gamma$ couplings through $W^+W^-Z$ production in $\gamma\gamma$
collisions}, J. Phys. G: Nucl. Part. Phys. \textbf{36}, 075007
(2009).
%
\bibitem{Koksal:2014}
M.~K\"{o}ksal, \emph{Anomalous quartic $ZZ\gamma\gamma$ couplings at
the CLIC}, Eur. Phys. J. Plus \textbf{130}, Article number: 75
(2015).
%
\bibitem{Koksal:2016}
M.~K\"{o}ksal, V.~Ari and A.~Senol, \emph{Search for anomalous
quartic $ZZ\gamma\gamma$ couplings in photon-photon collisioms},
Advances in High Energy Physics \textbf{2016}, Article ID: 8672391
(2016).
%
\bibitem{Braun:2008}
H.~Braun \emph{et al.} (CLIC Study Team), \emph{CLIC 2008
parameters}, CERN-OPEN-2008-021, CLIC-NOTE-764.
%
\bibitem{Boland:2016}
M.J.~Boland \emph{et al.} (CLIC and CLICdp Collaborations),
\emph{Updated baseline for a staged Compact Linear Collider},
CERN-2016-004 (arXiv:1608.07537).
%
\bibitem{Dannheim:2012}
D.~Dannheim \emph{et al.}, \emph{CLIC $e^+e^-$ Linear collider
studies}, in Proceedings of 2013 Community Summer Study on the
Future of U.S. Particle Physics: Snowmass on the Mississippi
(CSS2013), 29 July -- 6 August, 2013, MN, US (arXiv:1208.1402).
%
\bibitem{CLIC_BSM}
\emph{The CLIC Potential for New Physics}, eds. J.~de~Blas \emph{et
al.}, CERN Yellow report: Monographs, Vol.~3/2018, CERN-2018-009-M
(CERN, Geneva, 2018).
%
\bibitem{Franceschini:2020}
R.~Franceschini, \emph{Beyond the Standard Model physics at CLIC},
Int. J. Mod. Phys. A \textbf{35}, 2041015 (2020).
%
\bibitem{ATLAS_ions}
M.~Aaboud \emph{et al.} (ATLAS Collaboration), \emph{Evidence for
light-by-light scattering in heavy-ion collisions with the ATLAS
detector at the LHC}, Nat. Phys. \textbf{13}, 852 (2017); G.~Aad
\emph{et al.} (ATLAS Collaboration), \emph{Observation of
light-by-light scattering in ultraperipheral Pb + Pb collisions with
the ATLAS detector}, Phys. Rev. Lett. \textbf{123}, 052001 (2019).
%
\bibitem{CMS_ions}
D.~d'Enterria \emph{et al.} (CMS Collaboration), \emph{Evidence for
light-by-light scattering in ultraperipheral PbPb collisions at
$\sqrt{s} = $5.02 TeV}, Nucl. Phys. A \textbf{982}, 791 (2019).
%
\bibitem{Atag:2010}
S.~Ata\u{g}, S.C.~\.{I}nan and \.{I}.~\c{S}ahin, \emph{Extra
dimensions in $\gamma\gamma\rightarrow\gamma\gamma$ process at the
CERN-LHC}, JHEP \textbf{09}, 042 (2010).
%
\bibitem{Inan:2019}
S.C.~\.{I}nan and A.V.~Kisselev, \emph{Probe of the
Randall-Sundrum-like model with the small curvature via
light-by-light scattering at the LHC}, Phys. Rev. D \textbf{100},
095004 (2019).
%
\bibitem{Inan:2020_1}
S.C.~\.{I}nan and A.V.~Kisselev, \emph{A search for axion-like
particles in light-by-light scattering at the CLIC}, JHEP
\textbf{06}, 183 (2020).
%
\bibitem{Inan:2020_2}
S.C.~\.{I}nan and A.V.~Kisselev, \emph{Polarized light-by-light
scattering at the CLIC induced by axion-like-particles}, Chin. Phys.
C \textbf{45}, 043109 (2021).
%
\bibitem{Ginzburg:1981}
I.F.~Ginzburg, G.L.~Kotkin, V.G.~Serbo and V.I.~Telnov,
\emph{Colliding $\gamma e$ and $\gamma\gamma$ beams based on the
single-pass $e^+e^-$ colliders (of VLEPP Type)}, Nucl. Instrum.
Meth. \textbf{205}, 47 (1983); I.F.~Ginzburg, G.L.~Kotkin,
S.L.~Panfil, V.G.~Serbo and V.I.~Telnov, \emph{Colliding $\gamma e$
and $\gamma\gamma$ beams based on single-pass $e^+e^-$ accelerators
II. Polarization effects, monochromatization improvement},
\emph{ibid.} \textbf{219}, 5 (1984).
%
\bibitem{Cakir:2008}
O.~\c{C}akir, K.O.~Ozansoy, \emph{Unparticle searches through
gamma-gamma scattering}, Eur. Phys. J. C \textbf{56}, 279 (2008).
%
\bibitem{polarized_beams}
G. A. Moortgat-Pick \emph{et al.}, \emph{The role of polarised
positrons and electrons in revealing fundamental interactions at the
Linear Collider}, Phys. Rep. \textbf{460}, 131 (2008).
%
\bibitem{CLIC_lum}
R.~Franceschini, P.~Roloff, U.~Schnoor and A.~Wulzer, \emph{The
Compact Linear $e^+e^-$ Collider (CLIC): Physics Potential},
arXiv:1812.07986.
%
\bibitem{Almeida:2020}
E.~da Silva Almeida, O.J.P.~\'{E}boly and M.C.~Gonzalez-Garcia,
\emph{Unitarity constraints on anomalous quartic couplings}, Phys.
Rev. D \textbf{101}, 113003 (2020).
%
\bibitem{Zhang:2020}
Yan-Ju Zhang and Jie-Fen Shen, \emph{Probing anomalous $tqh$
couplings via single top production in associated with the Higgs
boson at the HE-LHC and FCC-hh}, Eur. Phys. J. C \textbf{80}, 811
(2020).
%
\bibitem{Enterria:2013}
D.~d'Enterria and G.G.~da~Silveira, \emph{Observing light-by-light
scattering at the Large Hadron Collider}, Phys. Rev. Lett.
\textbf{111}, 080405 (2013); Erratum, Phys. Rev. Lett. \textbf{116},
129901 (2016).
%
\bibitem{Fichet:2014_2}
S.~Fichet \emph{et al.}, \emph{Probing new physics in diphoton
production with proton tagging at the Large Hadron Collider}, Phys.
Rev. D \textbf{89}, 114004 (2014).
%
\bibitem{Fichet:2015}
S.~Fichet \emph{et al.}, \emph{Light-by-light scattering with intact
protons at the LHC: from standard model to new physics}, JHEP
\textbf{02}, 165 (2015).
%
\end{thebibliography}
\end{document}